\documentclass[reprint,aps,amsmath,amssymb,superscriptaddress]{revtex4-1}

\usepackage[utf8]{inputenc}
\usepackage{graphicx}
\usepackage{bm}
\usepackage{xcolor}
\usepackage[normalem]{ulem}
\usepackage[ruled,lined]{algorithm2e}
\usepackage{setspace}
\usepackage{comment}
\usepackage[a4paper,margin=1in]{geometry}
\usepackage{amsmath,amssymb,amsfonts,amsthm}
\usepackage{graphicx}
\usepackage{siunitx}
\usepackage{microtype}
\usepackage{url}
\usepackage[breaklinks]{hyperref}

\usepackage{geometry}
\definecolor{crimson}{HTML}{DC143C}
\definecolor{perfect_green}{HTML}{4FBF26}

\begin{document}

\title{Not all Chess960 positions are equally complex}

\author{Marc Barthelemy}
\email{marc.barthelemy@ipht.fr}
\affiliation{Universit\'e Paris-Saclay, CNRS, CEA, Institut de Physique Th\'eorique,
91191 Gif-sur-Yvette, France}
\affiliation{Centre d'Analyse et de Math\'ematique Sociales (CNRS/EHESS), Paris, France}
\affiliation{Complexity Science Hub, Vienna, Austria}

\begin{abstract}

We analyze strategic complexity across all 960 Chess960 (Fischer Random Chess) starting positions. Stockfish evaluations reveal a near-universal first-move advantage for White ($\langle E \rangle = +0.33 \pm 0.12$ pawns), indicating that the initiative is a robust structural feature of the game. To quantify decision difficulty, we introduce an information-based measure $S(n)$ that captures the cumulative information required to identify optimal moves over the first $n$ plies. This measure decomposes into White and Black contributions, $S_W$ and $S_B$, defining a total opening complexity $S_{\mathrm{tot}} = S_W + S_B$ and a decision asymmetry $A = S_B - S_W$. Across the ensemble, $S_{\mathrm{tot}}$ ranges from $2.6$ to $17.2$ bits, while $A$ spans from $-4.5$ to $+4.2$ bits (mean $\langle A \rangle = -0.26$), showing that openings are nearly evenly split between those that burden White and those that burden Black, with a slight average excess complexity for White. Standard chess (position \#518, \texttt{RNBQKBNR}) exhibits near-average total complexity and  asymmetry, yet lies far from the configuration that jointly minimizes evaluation imbalance and decision asymmetry. These results reveal a highly heterogeneous Chess960 landscape in which small rearrangements of back-rank pieces can substantially alter strategic depth and competitive balance. The classical starting position—despite centuries of refinement—appears not as an extremum, but as one configuration among many in a broad statistical ensemble.

\end{abstract}

\maketitle

\section{Introduction}

Chess, one of humanity's oldest and most studied strategic games, served as a simple testing ground for theories of decision-making, artificial intelligence, and complex systems \cite{Levy1988ComputerChessCompendium}. From a physics perspective, chess represents an ideal model system for studying decision-making complexity and information dynamics: it is fully deterministic, has well-defined rules, generates vast empirical datasets, and admits precise computational analysis of optimal strategies. The modern rules of chess crystallized in the 15th century, with the classical starting position---\texttt{RNBQKBNR} (Rook-Knight-Bishop-Queen-King-Bishop-Knight-Rook)---becoming the universal standard. This position was not derived from mathematical principles but emerged through cultural evolution and practical play over centuries~\cite{murray1913history}.

The dominance of opening theory in modern chess has led to a paradoxical situation: at the highest levels of play, the first 15--20 moves often follow extensively analyzed variations stored in databases containing millions of games. This `opening book' knowledge can extend so deeply that the creative, analytical phase of the game is delayed until the middlegame. The tension between memorization and genuine understanding has long concerned chess theorists. As early as 1792, the Dutch chess enthusiast Philip Julius van Zuylen van Nijevelt proposed randomizing starting positions to restore creative play~\cite{zuylen1792}. Two centuries later, World Champion Bobby Fischer recognized the same issue as a potential threat to the game's intellectual vitality~\cite{fide_chess960_history}. In 1996, Fischer proposed a variant that preserves chess's strategic depth while eliminating memorization advantages: Fischer Random Chess, developed in collaboration with Susan Polgar~\cite{fide_chess960_history}. Now standardized as Chess960 or `Fischer random', and more recently promoted as `Freestyle Chess'~\cite{freestyle}, this variant shuffles the pieces on the back rank subject to three constraints: (i) the two bishops must occupy opposite-colored squares, (ii) the king must be positioned between the two rooks to preserve castling rights, and (iii) White and Black must share identical mirrored arrangements. These rules generate exactly $960$ legal starting positions (see Appendix A), with the classical arrangement \texttt{RNBQKBNR} corresponding to index~\#518 in the standard numbering.

Chess960 gained official recognition when FIDE incorporated it into the Laws of Chess in 2008 and began sanctioning world championships in 2019~\cite{fide_chess960_history}. More recently, it received even more attention with the new `Freestyle Chess Grand Slam Tour' \cite{freestyle}. From the standpoint of statistical physics, Chess960 offers a unique opportunity: a discrete ensemble of 960 initial conditions evolving under identical dynamical rules, enabling systematic comparison of complexity across configurations. While Fischer's motivation was practical---restoring the importance of understanding over memorization---his innovation raises questions amenable to quantitative analysis: Are all 960 positions equally complex? Does the classical position occupy a special place in the complexity landscape? Can we quantify the `decision-making difficulty' intrinsic to each starting configuration? These questions parallel fundamental inquiries in statistical physics concerning how initial conditions influence the complexity of dynamical trajectories.

Recent advances in computer chess, particularly the development of engines like Stockfish capable of evaluating positions with superhuman accuracy~\cite{stockfish2024}, enable us to address these questions rigorously. The availability of large databases has enabled the emergence of a quantitative `science of chess,' with studies addressing innovation dynamics~\cite{Perotti2013InnovationChess}, statistical regularities in opening frequencies~\cite{Blasius2009ZipfOpenings}, network properties of move sequences~\cite{Maslov2009PowerLawsChess,farren2013analysis,Ribeiro2013MoveByMove,barthelemy2025fragility,cerioli2025ai}, memory effects and temporal correlations~\cite{Schaigorodsky2014MemoryCorrelations,Schaigorodsky2016MemoryEffects}, game complexity measures~\cite{Atashpendar2016SequencingChess,DeMarzo2023ComplexityOpenings,Chacoma2024EmergentComplexity}, human performance and cognition~\cite{Chowdhary2023QuantifyingPerformance,Chassy2011SequenceKnowledge,Sigman2010RapidChessRT}, and spatial control patterns~\cite{Barthelemy2023SpaceControl,barthelemy:2025}. These works demonstrate that chess provides a rich empirical system for studying decision-making under uncertainty.

In this work, we characterize the 960 initial positions using simple quantitative measures, and we introduce an information–cost measure $S(n)$ that captures the cumulative information required to identify optimal moves over the first $n$ plies of gameplay. We apply this measure systematically to the 960 Chess960 positions, including the classical position \#518, to construct an empirical `complexity landscape' of opening configurations. Our approach treats chess as a complex system where each move represents a decision under uncertainty, with $S(n)$ capturing the difficulty of this decision-making process. This framework allows us to investigate whether the classical starting position---selected by historical accident rather than optimization---occupies any privileged location in the space of possible chess games, and to identify configurations that may offer greater symmetry or fairness between White and Black.

\section{Results}

In this section, we present three complementary indicators characterizing the 960 starting positions: (i) the positional probability distribution of pieces in the initial configuration, (ii) the engine-based structural evaluation of all positions, and (iii) a measure of the complexity burden borne by White and Black during the first moves of the game.

While many other metrics could in principle be considered, our objective here is to focus on quantities directly related to structural imbalance and early-game decision difficulty. Additional measures—such as the branching factor—are briefly discussed in Appendix~B, where we explain why they are less central to the questions addressed in this study.

\subsection{Location probability}

\begin{figure}
  \centering
  \includegraphics[width=1.0\linewidth]{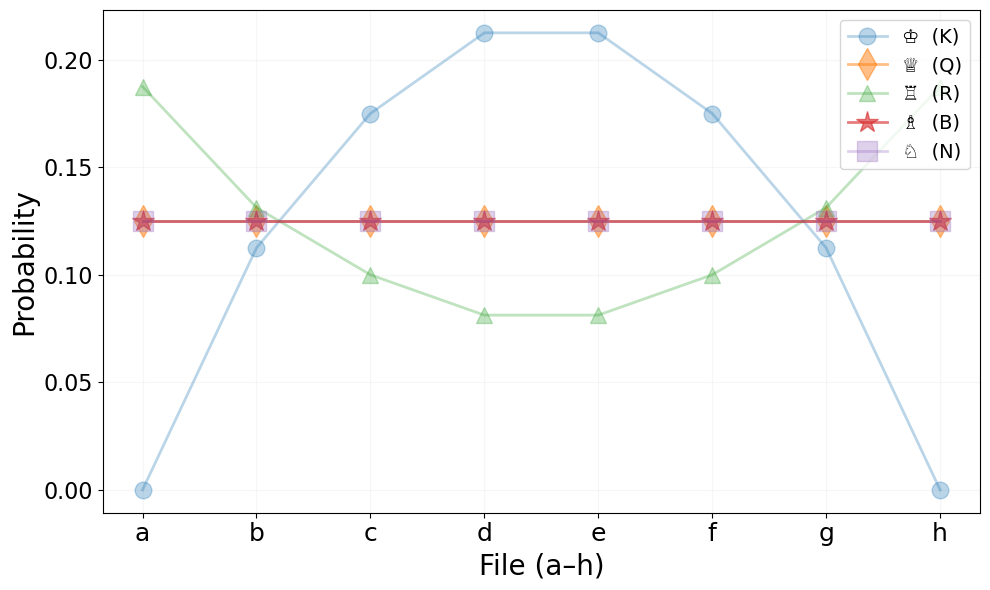}
  \caption{Probability distribution of pieces for the 960 initial positions across back-rank squares (from a to h). The bishops, knights, and queen each occupy any of the eight back-rank squares with equal probability $1/8$. By contrast, the king is more likely to appear near the central files, while the rooks are correspondingly more likely to be placed toward the sides.}
  \label{fig:proba}
\end{figure}
The 960 possibilities result from the following constraints (see Appendix A): The two bishops must occupy opposite-colored squares (one on a light square, one on a dark square); the king must be positioned between the two rooks, preserving the ability to castle kingside and queenside. Finally, Black pieces mirror the White arrangement, ensuring a priori fairness. These 960 positions are numbered from 0 to 959 and the standard chess position (\texttt{RNBQKBNR}) corresponds to the number 518 in this numbering \cite{initial_positions}. The simplest question we can ask under these constraints is: what is the probability that a given piece occupies a particular file? We show this probability in Fig.~\ref{fig:proba}. The bishops, knights, and the queen have equal probability across all eight squares ($1/8 = 12.5\%$). Due to castling constraints, the king and the rooks have non-uniform probability distributions across the eight squares. The king cannot be on the a or h file and we observe that it is most likely close to the central files (d-e), and naturally, the rooks have a larger probability to be on the sides. Fig.~\ref{fig:proba} shows the probability distribution for each piece type across the eight squares, revealing the symmetry and constraints of the Chess960 rules.

\subsection{Initial Stockfish evaluation}

In order to quantify the intrinsic structural asymmetry of Chess960 starting positions, we perform a systematic engine evaluation of all 960 configurations. The goal is not to model human perception or practical playing strength, but to estimate the underlying structural advantage encoded in each initial arrangement under near-optimal continuation.
To this end, we evaluated all 960 starting positions using Stockfish 17.1~\cite{stockfish2024} at fixed depth 30. The depth-30 score is interpreted here as an engine-level proxy for structural advantage, reflecting the evaluation after a sufficiently deep search while remaining computationally tractable. All computations were performed with NNUE enabled, using a single search thread (\texttt{Threads=1}) and a 1,GB hash table (\texttt{Hash=1024}). Each position was analyzed under a strict depth constraint without time limits, ensuring that every evaluation reached the prescribed search depth exactly.

\begin{figure}
\centering
\includegraphics[width=0.50\textwidth]{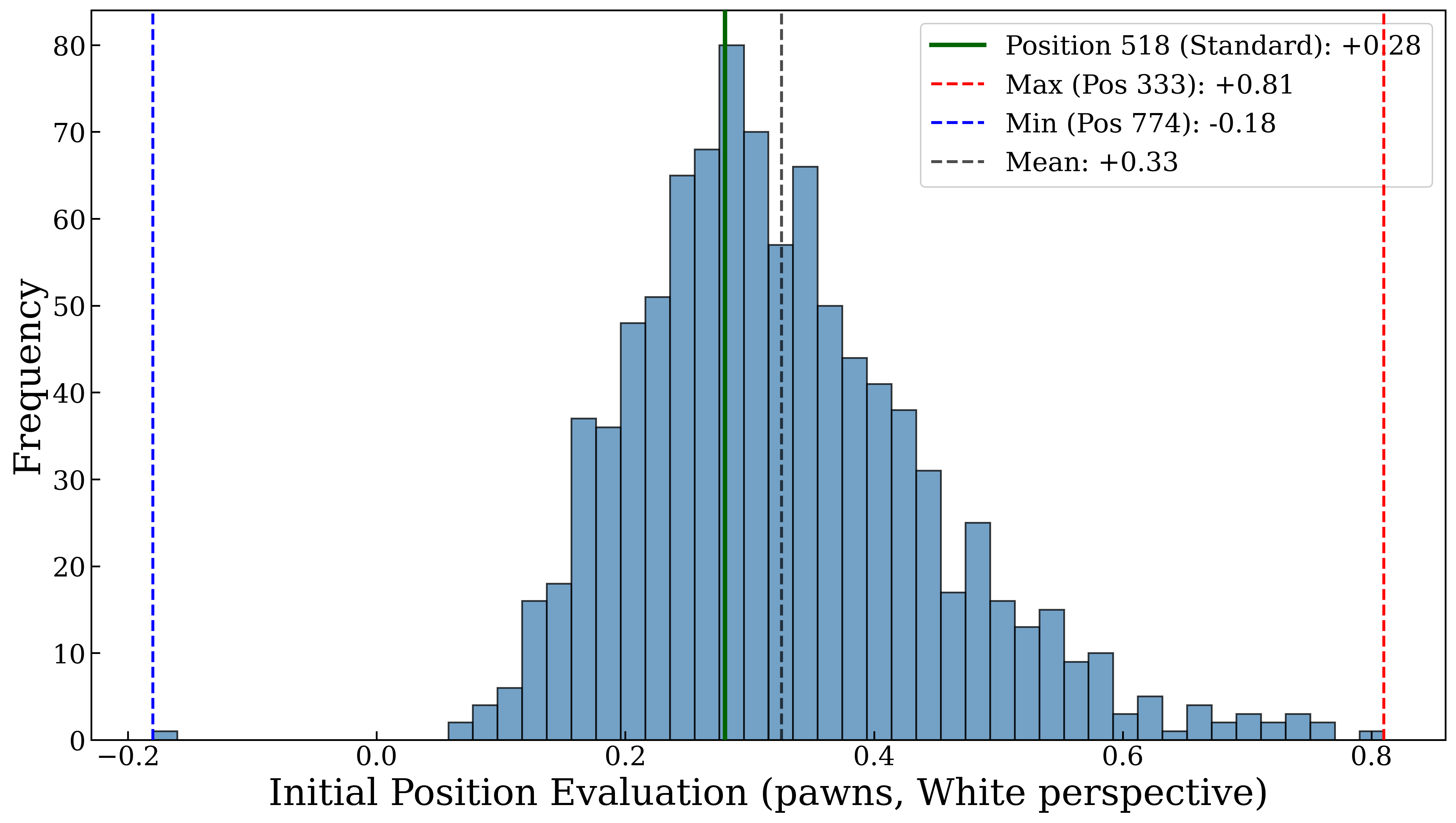}
\caption{Distribution of Stockfish initial evaluations across all 960 Chess960 starting positions (Stockfish 17.1, depth 30, \texttt{Threads=1}). The distribution is centered at $\langle E \rangle = +0.33 \pm 0.12$ pawns, with $99.9\%$ of positions favoring White. Vertical lines indicate position \#518 (standard chess, solid green). At this depth, position \#333 exhibits the maximum White advantage (red dashed), while position \#774 is the closest to perfect evaluation balance (blue dashed). The ensemble mean is shown as a black dashed line.}
\label{fig:initial_sf}
\end{figure}

The resulting distribution (Fig.~\ref{fig:initial_sf}) reveals a striking uniformity: the mean initial evaluation is $\langle E \rangle = +0.33 \pm 0.12$ pawns, with 959 of 960 positions (99.9\%) exhibiting positive evaluation favoring White. Only one position (\#774) achieves a small avantage for black ($E=-0.18$). This near-universal first-move advantage confirms that the privilege of moving first confers a measurable strategic benefit independent of piece configuration.

Standard chess (position \#518, \texttt{RNBQKBNR}) exhibits an initial evaluation of $+0.28$ pawns, placing it in the 37th percentile. This demonstrates that the classical starting position is statistically typical within the Chess960 landscape, neither amplifying nor diminishing the structural benefit of initiative.
The distribution's relatively narrow standard deviation ($\sigma = 0.12$ pawns) indicates remarkable consistency across disparate piece arrangements, suggesting that White's first-move advantage is a robust structural feature largely independent of the specific configuration. 

The most White-favorable position (\#333, \texttt{NRQBKRBN}) reaches the surprisingly large value of $+0.81$ pawns—approximately 2.5 times the mean. Inspection of the principal variation reveals that White can rapidly mobilize its pieces with the plan \texttt{Nb3}, \texttt{f4}, and \texttt{Bc5}, seizing central and kingside space while targeting the weakened dark squares around Black's king. The structural origin of this
asymmetry might lie in the initial placement itself: White's queen on \texttt{c1} and bishop on \texttt{d1} are already aligned along natural attacking diagonals, while the knight on \texttt{a1} finds an
efficient route to \texttt{b3} where it supports both the center and the \texttt{c5} outpost. More analysis of the variations are however necessary to inspect in more details this asymmetry. Another interesting case is the only position \#774 (QRBKNBRN) that gives an advantage to black, although very small ($-0.18$). The rarity of balanced configurations underscores that neutralizing White's first-move advantage requires a precise alignment of spatial and tactical factors.

From a game-theoretic perspective, the universality of White's advantage across Chess960 positions suggests that the first-move privilege is not an artifact of centuries of opening theory refinement in standard chess, but rather an intrinsic property of the game's mechanics. The advantage magnitude of $\approx 0.3$ pawns—roughly one-third of a minor piece—is consistent with empirical win-rate statistics from high-level play, where White scores a little above 50\% in standard chess. Fischer's motivation for Chess960 was to eliminate memorized opening theory, not to eliminate the first-move advantage itself, and our results confirm that this asymmetry persists as a fundamental feature of the game independent of initial configuration. This finding has implications for tournament fairness: while Chess960 successfully neutralizes preparation advantages, it preserves the need for color-balanced pairings or double-game formats to ensure equity between opponents.

\subsection{Effect of depth: instability of extremes}

\begin{figure}
\centering
\includegraphics[width=0.50\textwidth]{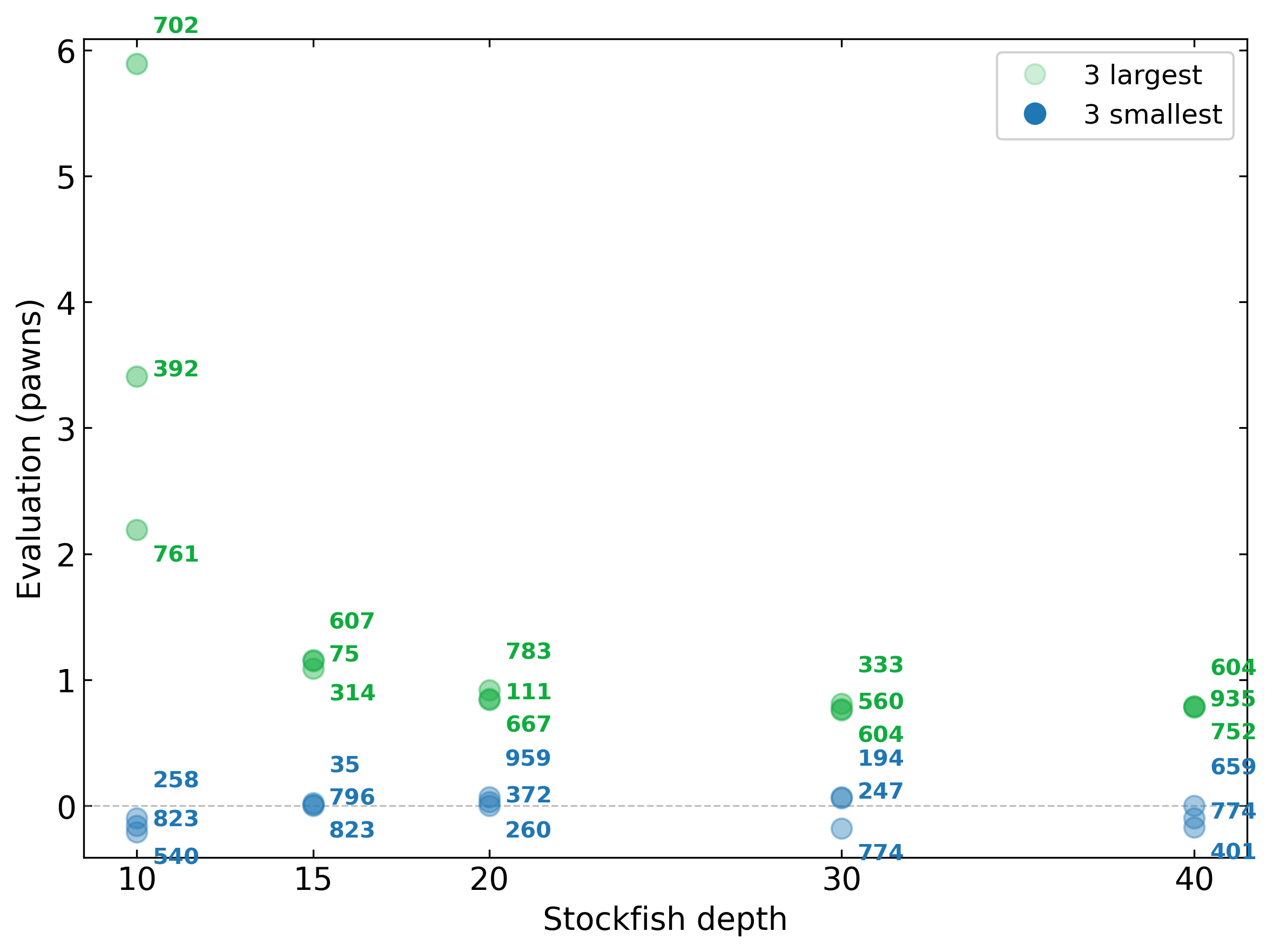}
\includegraphics[width=0.50\textwidth]{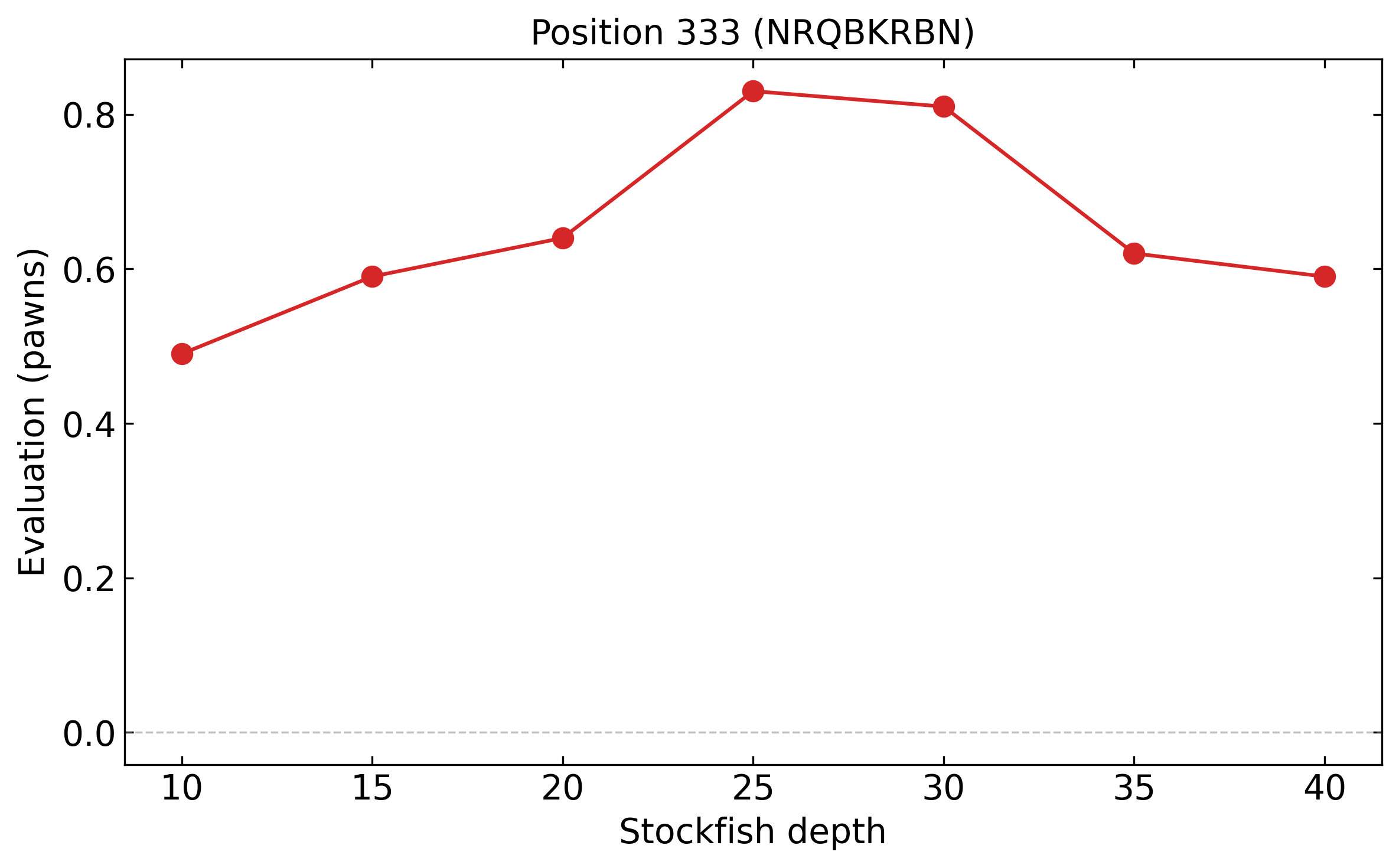}
\caption{(Top) Evaluation of the three highest- and three lowest-ranked initial positions as a function of search depth. (Bottom) Evaluation of position \#333 as a function of depth. All evaluations are expressed in pawns from White’s perspective.}
\label{fig:extremes_vs_depth}
\end{figure}

The outliers identified in the previous section were obtained at depth $30$. A natural question is whether these extreme evaluations remain stable as the search depth varies. In Fig.~\ref{fig:extremes_vs_depth}, we plot the three positions with the largest evaluations and the three with the smallest evaluations (as previously computed with Stockfish~17.1, \texttt{Threads=1}, NNUE enabled, identical hardware and hash settings) as a function of search depth.
The result is clear: extreme positions are not depth-stable. While aggregate statistics (mean, variance, overall distribution shape) remain robust, individual outliers fluctuate substantially, as illustrated in Fig.~\ref{fig:extremes_vs_depth}(top). At shallow depth the instability is striking. At depth 10, the top configuration (position 702) reaches an evaluation of $+5.89$ pawns, followed by $+3.41$ and $+2.19$. These extreme values completely disappear at greater depth. Already at depth 15, the top evaluations drop to $\approx +1.16$, and by depth 30 the maximum is $+0.81$. At depth 40, the leading positions cluster tightly around $+0.79$. Thus, the apparent large advantages observed at depth 10 are largely artifacts of shallow search. For bottom positions variations are much smaller. At depth 10, the lowest evaluations are only mildly negative ($-0.21$, $-0.16$, $-0.10$), but the identities of these configurations change repeatedly with depth. For instance, position 774 appears among the bottom three at depths 30 and 40, while other initially weak positions disappear from the extreme tail. Moreover, several configurations that are slightly negative at shallow depth become approximately neutral at intermediate depths, before stabilizing again.

The turnover in the ranking is equally informative. None of the top three positions at depth 10 remains in the top three at depth 20 or beyond. Conversely, position \#333—identified as the strongest configuration at depth 30 ($+0.81$)—does not appear among the leaders at shallow depth. Even between depths 30 and 40, although the numerical evaluations become closer, the composition of the top group still changes. To illustrate this sensitivity, we focus specifically on position \#333 and display in Fig.~\ref{fig:extremes_vs_depth} its evaluation as a function of search depth. The figure confirms that the evaluation is not stable with depth and does not exhibit clear convergence within the explored range. While the advantage peaks around depths 25–30, it subsequently decreases, reaching approximately $+0.5$ pawns at depth 40.

This behavior indicates that extreme evaluations are particularly sensitive to deeper tactical or strategic resources revealed by extended search. Positions that appear highly favorable at shallow depth may rely on lines that are later mitigated by longer forcing sequences or defensive resources uncovered at greater depth. More generally, this raises two questions: (i) do some positions converge rapidly and remain stable across depths, while others exhibit strong depth sensitivity? and (ii) which structural features control this stability? Addressing these issues would require a systematic study and the introduction of a quantitative measure of evaluation stability, allowing positions to be classified according to their sensitivity to search-horizon effects.

In summary, while global evaluation statistics are robust, the ranking of extreme Chess960 starting positions is not invariant with respect to search depth. Claims about the “most advantageous” or “most disadvantageous” configurations should therefore be supported by an explicit depth-stability analysis. Search depth should be interpreted as defining the resolution at which the decision landscape is probed.

\subsection{Characterizing the opening complexity}


The measures discussed so far, however, do not quantify how difficult it is to play a given position as White or Black, nor do they assess whether one starting configuration is intrinsically more complex than another. To address these questions, we introduce an information-based measure of decision complexity that captures the difficulty of identifying the best moves.

Consider a position where a chess engine evaluates (in centipawns) the best move with score $E_1$  and the second-best move with score $E_2$. The difference $\Delta = E_1 - E_2$ quantifies how much `better' the optimal move is~\cite{Barthelemy2023SpaceControl,barthelemy:2025}. For a player with discrimination ability $\Delta_0$ (the minimum evaluation difference they can reliably perceive), we model the probability $P(\text{optimal})$ of choosing the optimal move (over the second best move) using a Boltzmann-like factor (also called the softmax function)
\begin{equation}
P(\text{optimal}) = \frac{1}{1 + e^{-\Delta/\Delta_0}}
\end{equation}
This form has the expected limiting behavior: when $\Delta \gg \Delta_0$, $P \to 1$ (the best move is obvious), and when $\Delta \ll \Delta_0$, $P \to 1/2$ (almost indistinguishable moves). To quantify the difficulty of identifying the optimal move at a given ply, we associate to each position an information content or cost (in bits) given by
\begin{equation}
S(\Delta) = -\log_2 P= \log_2\!\left(1 + e^{-\Delta/\Delta_0}\right),
\label{eq:SDelta}
\end{equation}
where $\Delta$ is the evaluation gap between the best and second-best moves, and $\Delta_0$ sets the discrimination scale (analogous to a noise or `temperature' parameter). Importantly, $S(\Delta)$ is \emph{not} an entropy of a probability distribution. Instead, it measures the amount of information required to resolve the local ambiguity in the game tree. The form~\eqref{eq:SDelta} is identical to the log-partition function of a two-state softmax (logit) choice model, widely used in decision theory and statistical physics. In that context, $S(\Delta)$ plays the role of the `cost' of discriminating between nearly equivalent alternatives. This interpretation yields intuitive limits:  
\begin{itemize}
    \item If $\Delta \ll \Delta_0$, the two moves are nearly indistinguishable: identifying the optimal one requires close to a full bit of information, $S(\Delta) \approx \log_2 2 = 1$ bit.
    \item If $\Delta \gg \Delta_0$, the best move is effectively forced: the information cost collapses to zero, $S(\Delta) \approx 0$ bit.
\end{itemize}

For a trajectory of $n$ plies, the cumulative information required to navigate the corresponding branch of the search tree is
\begin{equation}
S(n) = \sum_{i=1}^{n} S(\Delta_i)
= \sum_{i=1}^{n} \log_2\left(1 + e^{-\Delta_i/\Delta_0}\right),
\label{eq:Sn}
\end{equation}
where $\Delta_i$ denotes the evaluation gap at ply $i$. The average information per move, $S(n)/n$, characterizes the typical decision difficulty encountered along the trajectory. Thus, $S(n)$ provides an information-theoretic measure of how demanding it is to follow a given sequence of moves (for example, an opening line) under optimal play.
The quantity $S(n)$ has several natural properties. It is monotonically increasing with $n$, reflecting the cumulative nature of information along the line of play. Positions with large $\Delta_i$ (quasi-forced moves) contribute negligibly, whereas positions with small $\Delta_i$ dominate the sum. The overall scale of $S(n)$ depends on the discrimination parameter $\Delta_0$, which encodes the sensitivity to small evaluation differences. In Appendix~C, we further show that $S(10)$ correlates with the mean thinking time $\tau$ of human players: $\tau$ increases with $S(10)$, supporting its interpretation as a meaningful proxy for decision-making complexity.
Importantly, $S(n)$ is not intended to model human computational limits. Rather, it quantifies the intrinsic \emph{difficulty of identifying the optimal path} among competing continuations (here in the early game). In this sense, $S(n)$ is a structural property of the position: it measures how sharply the optimal move is separated from alternatives along the principal variation. It is therefore independent of player strength and instead characterizes the geometry of the decision landscape itself.

Estimating a precise value for $\Delta_0$ is non-trivial, and any choice is necessarily somewhat arbitrary. In Appendix~C, we show that the percentile ranking of a representative position (the standard position \#518) varies only weakly with $\Delta_0$, indicating robustness with respect to this parameter. We also provide empirical evidence that $\Delta_0 \approx 10$~cp corresponds to a natural discrimination scale at expert (Grandmaster) level. In what follows, we therefore focus on this high-skill regime and set $\Delta_0 = 10$~cp.
\begin{figure*}
  \centering
\includegraphics[width=\textwidth]{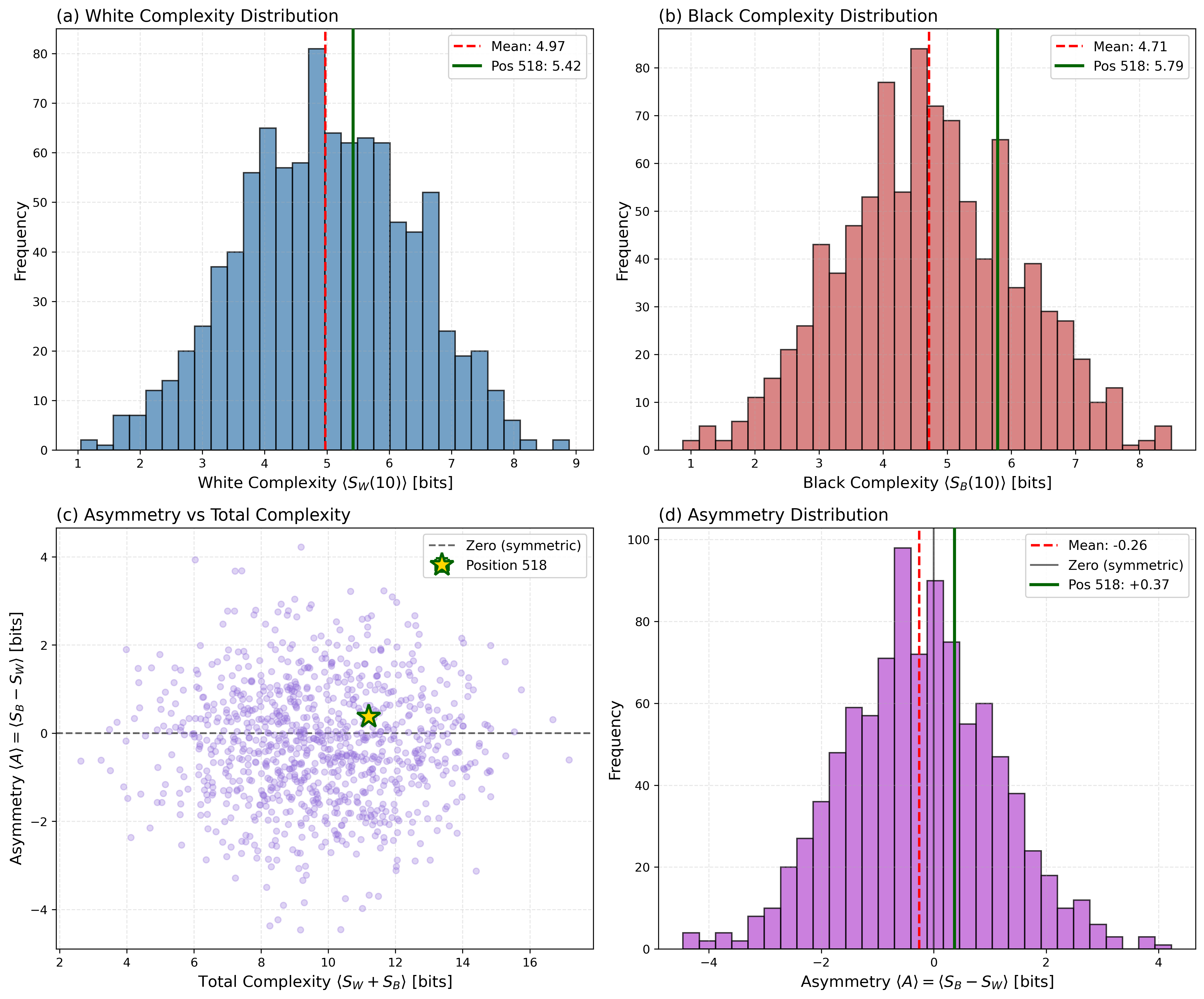}
\caption{Information cost analysis across the 960 Chess960 starting positions, computed for $n = 10$ plies per player under optimal play (Stockfish 17.1, depth 15, discrimination threshold $\Delta_0 = 10$~cp).
\textbf{(a)} Distribution of White's cumulative cost $S_W(10)$, quantifying the decision complexity faced by White across starting configurations. The green vertical line marks position \#518 (standard chess), with $\langle S_W \rangle = 5.42$ bits. 
\textbf{(b)} Distribution of Black's cumulative cost $S_B(10)$. Position \#518 yields $\langle S_B \rangle = 5.79$ bits, slightly above the ensemble mean. 
\textbf{(c)} Asymmetry $A = S_B - S_W$ versus total information cost $S_{\mathrm{tot}} = S_W + S_B$. Negative asymmetry indicates greater decision complexity for White; positive values indicate greater complexity for Black. The average asymmetry is slightly negative, $\langle A \rangle = -0.26$ bits (standard deviation $1.36$), while the average total cost is $\langle S_{\mathrm{tot}} \rangle = 9.69$ bits (standard deviation $2.4$). Standard chess (position \#518, gold star) exhibits positive asymmetry ($A_{518}= +0.37$ bits) and a relatively high total cost ($11.2$ bits, 72.5th percentile). 
\textbf{(d)} Distribution of asymmetry across all positions. The mean asymmetry is $\langle A \rangle = -0.26$ bits, indicating a slight average structural complexity advantage for Black across Chess960. Standard chess lies at the 69th percentile of the asymmetry distribution, placing it close to the ensemble average.}
\label{fig:asym}
\end{figure*}

The information cost can be decomposed by color, yielding $S_W(n)$ for White and $S_B(n)$ for Black over $n$ plies each. A natural measure of positional fairness is the symmetry between these quantities: in a balanced starting configuration, $S_W(n) \approx S_B(n)$, at least in the opening phase, indicating comparable decision-making difficulty for both players. We compute these quantities for all 960 starting positions over 10 plies (also called `half-moves') per player (20 total plies corresponding to a total of 10 moves).
Also, we restricted the analysis here to depth 15 because evaluating 10 moves for all $960$ games entails a substantial computational cost. Results for distribution of $S_B$ and $S_W$ are shown in Fig.~\ref{fig:asym}(a,b). The distributions of White and Black decision complexity, 
$P(S_W)$ and $P(S_B)$ are both approximately Gaussian with similar widths, indicating that most Chess960 positions impose a comparable level of opening difficulty for each player. Their means differ slightly—$\langle S_W\rangle\approx 4.97$ bits and $\langle S_B\rangle\approx 4.71$ bits —showing that White typically faces a marginally higher decision load during the first ten plies. The overall spread (roughly 6 bits) reflects substantial variability across starting positions, while the near-identical shapes of the two distributions suggest that the structural factors governing early complexity act symmetrically for both colors. Standard chess (\#518) sits near the centers of both distributions, confirming that its absolute opening difficulty is typical within the broader Chess960 ensemble.

The information cost framework reveals important variations in decision-making burden between White and Black across Chess960 positions. We define the asymmetry as 
\begin{align}
A = S_B(n) - S_W(n)
\end{align}
which quantifies the differential complexity: positive values indicate Black faces more difficult decisions, while negative values indicate White's burden is greater. Analysis of all 960 starting positions (Fig.~\ref{fig:asym}(d)) reveals a mean asymmetry of $\langle A \rangle = -0.26$ bits (and standard deviation $\pm 1.36$ bits), suggesting a weak but systematic bias wherein White generally faces slightly more complex decision trees. This negative mean may reflect the fundamental asymmetry inherent in chess: White moves first and must immediately confront the full complexity of opening choices, while Black's initial responses are partially constrained by White's play. The substantial variation from $-4.4$ to $+4.2$ bits) indicates high game-to-game variability, reflecting the sensitivity of decision complexity to openings for specific initial conditions. 

Standard chess (position \#518, \texttt{RNBQKBNR}) exhibits $\langle S_W \rangle = 5.42$ bits and $\langle S_B \rangle = 5.79$ bits, yielding asymmetry $\langle A \rangle = +0.36$ bits and total complexity $\langle S_{\mathrm{tot}} \rangle = 11.2$ bits. The positive asymmetry places standard chess at the 69th percentile, indicating that Black faces moderately harder decisions than in most Chess960 configurations. The total complexity percentile of 72.5th indicates that standard chess exhibits above average overall complexity, but can be considered as neither exceptionally simple nor complex relative to the broader Chess960 ensemble.

The scatter plot of asymmetry versus total information cost (Fig.~\ref{fig:asym}c) reveals no significant correlation between these quantities, indicating that overall game complexity is independent of the White-Black balance. Highly complex games can be either symmetric or asymmetric, and conversely, simple games show comparable variation in burden distribution. This independence suggests that Chess960's design successfully decouples two distinct dimensions of strategic depth: the total decision-making challenge and its allocation between players.

\subsection{Illustration 1: Searching for jointly balanced positions}

The two fairness metrics introduced above—initial evaluation $E$ and
decision asymmetry $A$—capture distinct aspects of competitive balance: $E$ measures the expected outcome advantage, while $A$ quantifies the imbalance in cognitive burden between players.
A natural question is whether these two dimensions are correlated.
Do positions favoring White in evaluation also impose greater
decision complexity on Black?

\begin{figure}
  \centering
  \includegraphics[width=1.0\linewidth]{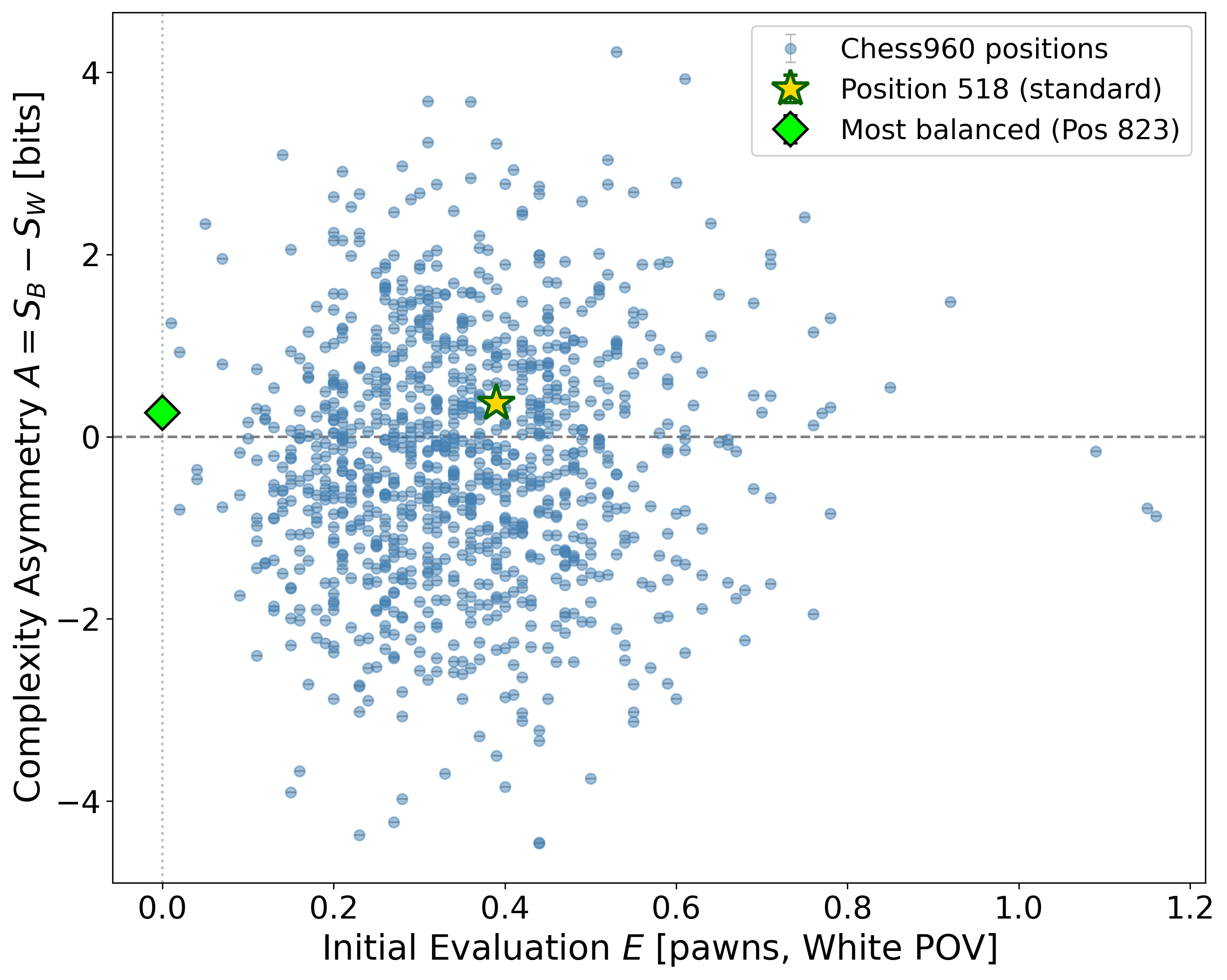}
  \caption{Complexity asymmetry $A = S_B - S_W$ versus initial Stockfish evaluation $E$
  (Stockfish 17.1, depth 15, single thread) for all 960 Chess960 starting positions.
  The weak correlation ($r \approx 0.15$) indicates near-independence
  between evaluation balance and decision symmetry.
  The gold star marks position \#518 (standard chess).
  The green diamond indicates the configuration that minimizes the
  normalized distance to the origin in the $(E,A)$ plane at this depth.}
  \label{fig:AvsE}
\end{figure}

Figure~\ref{fig:AvsE} plots the asymmetry
$A = S_B - S_W$ against the initial Stockfish evaluation $E$
(depth 15) for all 960 starting positions.
The correlation is weak ($r \approx 0.15$),
indicating that evaluation advantage and decision complexity
are largely independent characteristics of a position.
Positions that are evaluation-balanced do not necessarily yield
balanced cognitive demands, and vice versa.

This near-independence motivates the search for configurations that are simultaneously close to zero in both metrics.
We therefore define a joint balance score as the normalized distance
\begin{equation}
d = \sqrt{\left(\frac{E}{\sigma_E}\right)^2 +
          \left(\frac{A}{\sigma_A}\right)^2},
\end{equation}
where $\sigma_E$ and $\sigma_A$ are the ensemble standard deviations
of $E$ and $A$, respectively.
This provides a scale-invariant measure of proximity to the ideal
balanced point $(E,A)=(0,0)$.

At depth 15, this criterion identifies position \#823
(\texttt{RKBNQRNB}) as the closest configuration to the origin,
with $E \approx 0$ pawns and $A \approx +0.26$ bits.
Importantly, however, the identity of the minimizing position
is not universal: as discussed above, outliers and fine
rankings can depend on engine depth and evaluation settings.
The purpose of this construction is therefore not to single out
a unique ``optimal'' starting position, but rather to illustrate
how joint fairness criteria can be defined and systematically explored.

Standard chess (position \#518) lies at a moderate distance from the origin, and although standard chess is not extreme in evaluation,
it exhibits a non-negligible structural decision asymmetry.

A deeper chess-specific interpretation of jointly balanced configurations requires further analysis.
Relevant structural factors may include rook placement
(e.g., both rooks on the same flank), king position and
castling structure, early tactical constraints,
bishop geometry and diagonal accessibility,
and the availability of immediate pawn breaks.
Understanding how such geometric features influence
both $E$ and $A$ would provide a bridge between
statistical characterization and concrete opening theory.

\subsection{Illustration 2: the most complex positions} 

With this consolidated protocol (and depth $15$), the most complex Chess960 initial position is position $524$ (RBNQKNBR), with a total complexity $S_{\mathrm{tot}} = 17.17$ bits. For comparison, the standard chess starting position (position $518$, RNBQKBNR) has 
$S_{\mathrm{tot}} = 11.20$ bits (and initial evaluation $E = 39$ cp).

Interestingly, this most complex position remains very close to the standard chess configuration, differing only by the permutation of a few pieces (B and N on both king and queen sides). Its engine evaluation is also relatively small, $E = 24$ cp, which is even closer to equality than the standard chess position. This observation confirms that high combinatorial complexity is not associated with a large initial imbalance in the engine evaluation.

The other configurations with the largest complexity have also values far above that of the standard starting position. The five most complex positions identified in our analysis are: position $524$ (RBNQKNBR) with $S_{\mathrm{tot}}=17.166$ ($S_W=8.886$, $S_B=8.280$); position $444$ (RBNNQKBR) with $S_{\mathrm{tot}}=16.685$ ($S_W=8.190$, $S_B=8.495$); position $932$ (RBBKRNQN) with $S_{\mathrm{tot}}=15.746$ ($S_W=7.383$, $S_B=8.362$); position $535$ (RNBKQNRB) with $S_{\mathrm{tot}}=15.540$ ($S_W=7.756$, $S_B=7.784$); position $372$ (NBBRKRNQ) with $S_{\mathrm{tot}}=15.277$ ($S_W=7.896$, $S_B=7.381$).

It would be interesting to understand what these positions have in common and what governs their large complexity values. Finally, it is important to stress that these values are obtained at search depth $15$. While this depth already provides a reasonable proxy for human-level analysis, deeper searches may modify the ranking of the most complex positions, as discussed above. The configurations reported here should therefore be interpreted as belonging to a family of extremely high-complexity starting positions rather than as defining an absolutely fixed ordering across all possible search depths.

\section{Discussion and Conclusion}

We introduced an information–cost framework to quantify decision complexity across all 960 Chess960 starting positions. Two main results emerge. First, the Chess960 ensemble forms a highly heterogeneous landscape: total opening complexity and decision asymmetry vary widely, showing that small rearrangements of back-rank pieces can substantially alter both strategic depth and competitive fairness. In particular, the asymmetry $A = S_B - S_W$ spans several bits, with some positions disproportionately burdening White and others Black. A small random sample of starting positions does not, in general, guarantee competitive balance. Our framework provides a quantitative basis for designing more principled selection protocols in Freestyle tournaments.

Second, the classical starting position (position \#518) occupies a statistically unremarkable location in this landscape. It exhibits a typical evaluation advantage for White, moderate total complexity, and a slightly positive asymmetry indicating a greater cognitive burden for Black. Despite centuries of play, the traditional arrangement does not maximize complexity nor minimize imbalance. This suggests that its historical persistence may reflect aesthetic symmetry, pedagogical accessibility, or cultural contingency rather than structural optimality.

More broadly, this work demonstrates how tools from information theory and statistical physics can characterize strategic complexity in deterministic decision systems. The framework is not specific to chess: it applies whenever optimal actions can be evaluated and the informational effort required to distinguish them can be quantified. Extending this approach to other chess variants, historical games, or different strategic systems (Go, Shogi, and modern board games) would allow systematic comparisons of complexity landscapes and provide new insight into the evolution and design of competitive games.\\

\medskip
\begin{acknowledgments}
  {\bf Acknowledgments}\\
  I thank Giordano de Marzo for his suggestion about thinking time and complexity, and Richard Botley for interesting discussions. I also thank the GothamChess and Blitzstream channels for providing valuable food for thought about chess.
\end{acknowledgments}

\appendix

\section{Appendix A: 960 positions}

he number $960$ follows directly from elementary combinatorics. Labeling the back-rank squares (from a to h), the bishops can be placed on any pair of light and dark squares, giving $4 \times 4 = 16$ combinations. The queen may then occupy any of the remaining six squares. This leaves five squares to be filled by the multiset $\{R,K,R,N,N\}$. With 5 squares left, there are $\binom{5}{2} = 10$ ways to place the two knights; the remaining three squares must then be occupied by the two rooks and the king, with the king between the rooks, which yields a unique arrangement for these three pieces. Altogether, this gives $16 \times 6 \times 10 = 960$ possible starting positions. This deterministic construction also provides a canonical indexing scheme from $0$ to $959$, known as the Scharnagl enumeration algorithm~\cite{scharnagl2004fischer,WikipediaChess960Numbering}. In this scheme, each starting position is uniquely encoded by three integers: (i) the \emph{bishop code}, selecting one of the $16$ opposite-colored bishop pairs; (ii) the \emph{queen position}, selecting one of the $6$ remaining squares after the bishops are placed; and (iii) the \emph{N5N code}, indexing one of the admissible arrangements of the multiset $\{R,K,R,N,N\}$ on the final five squares under the requirement that the king lies between the two rooks. The mapping to an index is given by
$\mathrm{idn} \;=\; (\text{bishop code}) 
\;+\; 16 \times (\text{queen position}) 
\;+\; 96 \times (\text{N5N code})$. For the classical starting position \texttt{RNBQKBNR}, this yields $
\mathrm{idn} = 6 \;+\; 16 \times 2 \;+\; 96 \times 5 = 518$ which is the standard numbering for classical chess. This indexing provides a compact and reproducible method for scanning all $960$ starting configurations and guarantees a one-to-one correspondence between integers in $[0,959]$ and legal Chess960 starting positions.

\section{Appendix B: Branching factor}

To quantify the complexity evolution across different Chess960 starting positions, we analyze two key metrics: the branching factor $b(n)$ and the position space size $N(n)$ (i.e. the number of different positions) as functions of ply depth $n$. This analysis characterizes the raw combinatorial complexity of the game tree, independent of strategic considerations or optimal play. It quantifies the fundamental decision-making burden at each stage of the game: how many distinct game states a player must consider when planning ahead.

\begin{figure}
  \centering
  \includegraphics[width=1.0\linewidth]{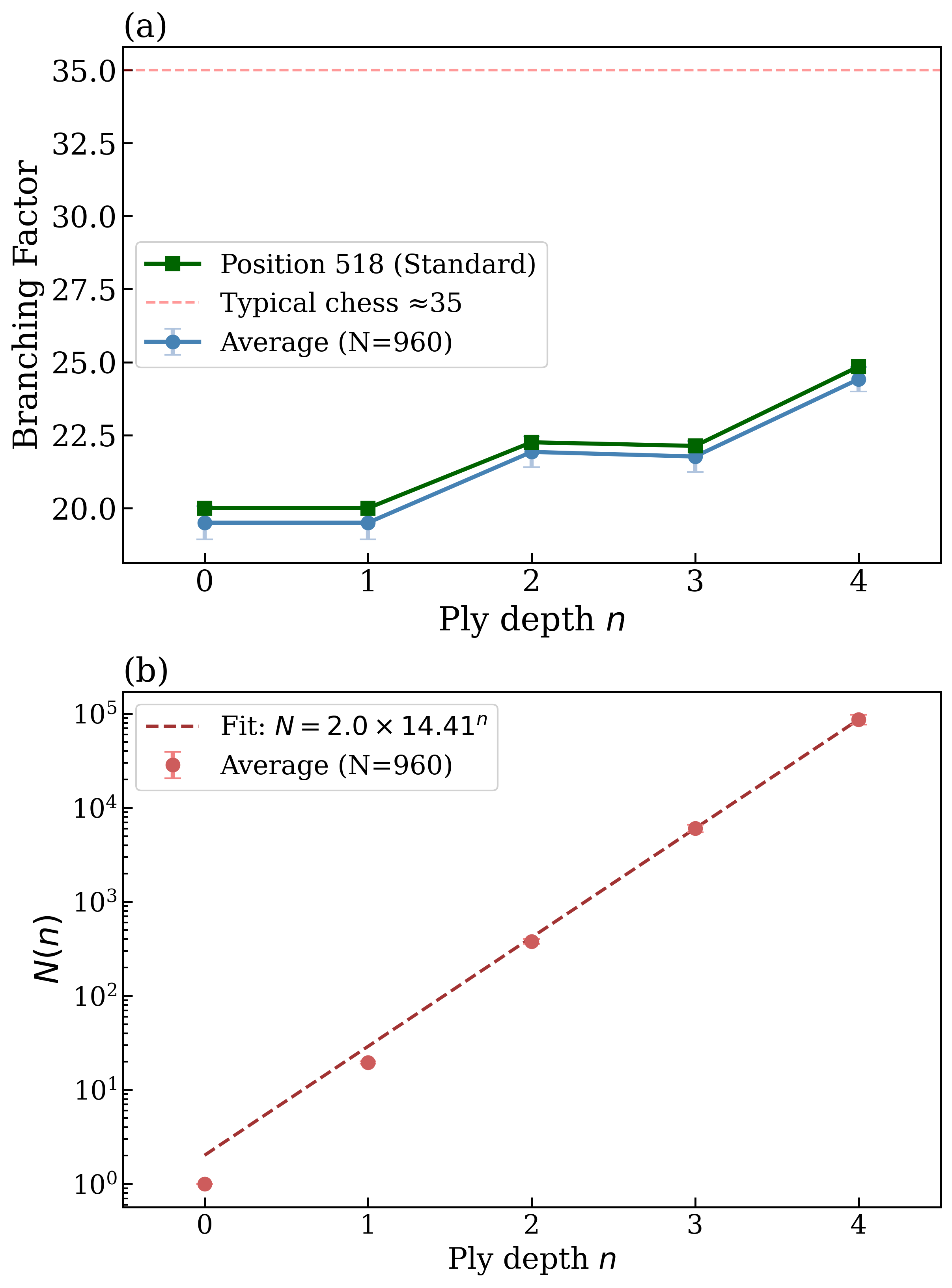}
  \caption{(a) Average branching factor $\langle b \rangle$ versus ply depth $n$ across all 960 starting configurations (blue), compared with standard chess (position \#518, green). Error bars indicate one standard deviation. The dashed line marks the commonly cited middlegame value $b \approx 35$. (b) Number of unique positions $N(n)$ versus ply depth on a semi-logarithmic scale, showing exponential growth $N(n) \sim b_{\mathrm{eff}}^n$ with effective branching factor $b_{\mathrm{eff}} \approx 14.41$, significantly smaller than $\langle b \rangle$ due to transpositions.}
  \label{fig:branching}
\end{figure}

For each position, we compute $N(n)$ defined as the number of unique legal positions reachable at ply depth $n$. We then compute the branching factor $b(n)$ defined as the ratio of consecutive position counts
\begin{equation}
        b(n) = \frac{N(n+1)}{N(n)}
\end{equation}
This ratio quantifies the average number of accessible successor states per position at depth $n$. Figure~\ref{fig:branching} presents a comparative analysis between position 518 (standard chess) and the ensemble average across all sampled positions. Panel (a) shows the evolution of the branching factor $b(n)$ versus ply depth. The average branching factor across sampled positions (blue curve with error bars representing one standard deviation) exhibits typical values of order $20$ consistent with previous estimates \cite{Barthelemy2023SpaceControl}. Note that branching factors around $b \sim 35$, often cited in computer chess literature, correspond to middlegame positions where piece development leads to increased move options~\cite{Barthelemy2023SpaceControl}. Position \#518 (green curve) closely tracks this average, with deviations remaining within one standard deviation throughout the analyzed range.

Panel (b) displays the growth of unique positions $N(n)$ on a semi-logarithmic scale. The approximately linear behavior confirms exponential growth, $N(n) \sim \exp(\lambda n)$. The extremely small dispersion around the average (smaller than the symbol size) indicates that all starting positions generate state spaces of essentially identical size. Note that the effective growth rate $\lambda$ is notably smaller than what would be predicted from the branching factor alone ($\lambda < \ln\langle b \rangle$). This reduction arises because different move sequences can converge to the same board configuration---a phenomenon known as transposition in chess terminology---causing the position space to grow more slowly than the space of legal move sequences.
From the perspective of branching factors and the number of unique positions, our results show no significant dependence on the initial arrangement; in this sense, all 960 starting positions are equivalent.

\section{Appendix C: $\Delta_0$ and time to make a move}

\subsection{Effect of $\Delta_0$}

The threshold parameter $\Delta_0$ controls the definition of a ``difficult'' move and therefore directly affects both the asymmetry and the total complexity. 
In principle, varying $\Delta_0$ should modify the absolute values of these quantities: 
a smaller $\Delta_0$ makes more moves count as difficult, whereas a larger $\Delta_0$ selects only the most significant evaluation differences. 
However, our results should not depend critically on the precise choice of $\Delta_0$. 
In particular, relative comparisons between starting positions are expected to remain stable provided $\Delta_0$ lies within a reasonable range.

\begin{figure}
\centering
\includegraphics[width=0.5\textwidth]{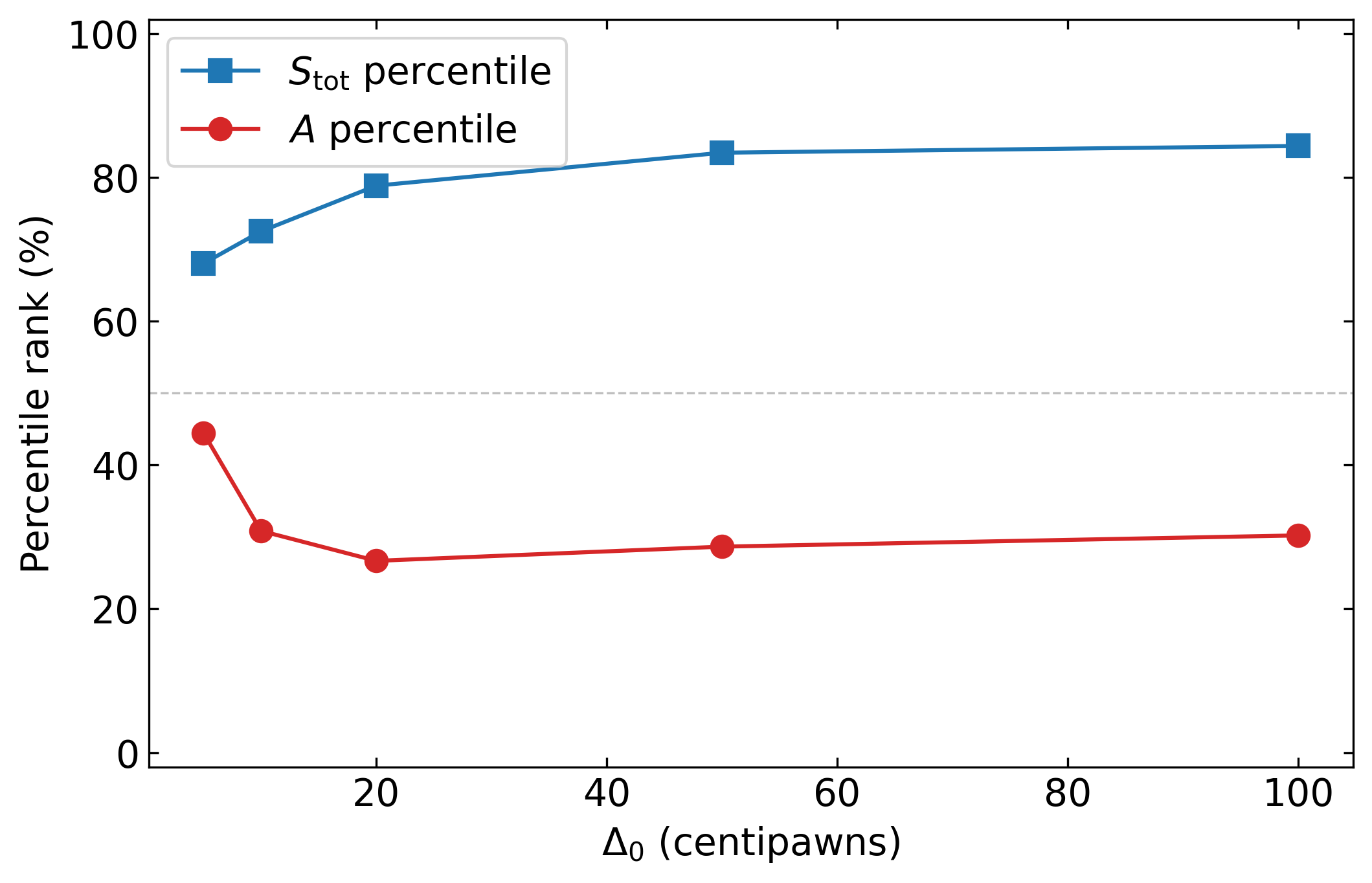}
\caption{Percentile rank of the total complexity $S_{\mathrm{tot}}$ and the asymmetry $A$ for position \#518 as a function of $\Delta_0$, at engine depth 15. The percentile varies smoothly with $\Delta_0$, 
indicating robustness with respect to the choice of threshold.}
\label{fig:percentile_vs_delta}
\end{figure}

To test this robustness, we computed the asymmetry and total complexity for all 960 starting positions as a function of $\Delta_0$, and evaluated the percentile rank of a representative position (the standard position \#518) as $\Delta_0$ varies. 
Figure~\ref{fig:percentile_vs_delta} shows the evolution of this percentile for engine depth $15$, which is the depth used for most computations in the main text to ensure consistency.

We observe that the percentile varies smoothly with $\Delta_0$, without abrupt transitions. 
For very small $\Delta_0$, almost all moves are classified as difficult; 
as a consequence, most positions become statistically similar and the percentile of position \#518 approaches $50\%$, as expected for a nearly homogeneous ensemble. 
In contrast, for $\Delta_0 \gtrsim 10$~cp, the percentile stabilizes and varies only weakly with further increases of the threshold.

These results show that while $\Delta_0$ influences absolute complexity values, 
the relative ranking of positions—and therefore our main conclusions—remain robust over a broad and physically meaningful range of thresholds. 
In particular, the choice $\Delta_0 = 10$~cp (used in the main text for expert-level analysis) lies well within this stable regime.

\subsection{Empirical analysis}

To assess whether the evaluation gap $\Delta$ serves as a meaningful proxy for decision-making complexity, we perform an empirical analysis of players’ thinking times using a large, publicly available database of online games played on Lichess, which includes move-level time stamps~\cite{LumbraGigabase}.

Two successive filters are applied to ensure comparability across games. First, we retain only moves played by players with a known Elo rating
$\geq 2000$, thereby restricting the analysis to expert-level play and reducing fluctuations associated with lower skill levels. 
Second, we use the PGN \texttt{TimeControl} header
(e.g.\ \texttt{5400+30}, \texttt{300+3}) to classify games by time format.
We focus here on the Blitz category ($60 \leq T_0 < 600$\,s,
corresponding to $1$--$10$ minutes), where time pressure is substantial
and thinking time is expected to reflect decision difficulty more directly.

For this time-control category, we compute the binned average thinking time
$\langle \tau \rangle$ as a function of the evaluation gap
\[
\Delta = E_1 - E_2,
\]
where $E_1$ and $E_2$ denote the engine evaluations (in centipawns)
of the best and second-best moves at depth $15$.
The $\Delta$ axis is partitioned into the following bins:
$0$--$5$, $5$--$10$, $10$--$20$, $20$--$50$, $50$--$100$,
$100$--$200$, $200$--$500$, and $500$--$1000$\,cp.
Bins containing fewer than $10$ moves are discarded.
Error bars represent the standard error of the mean.

\begin{figure}
\centering
\includegraphics[width=0.5\textwidth]{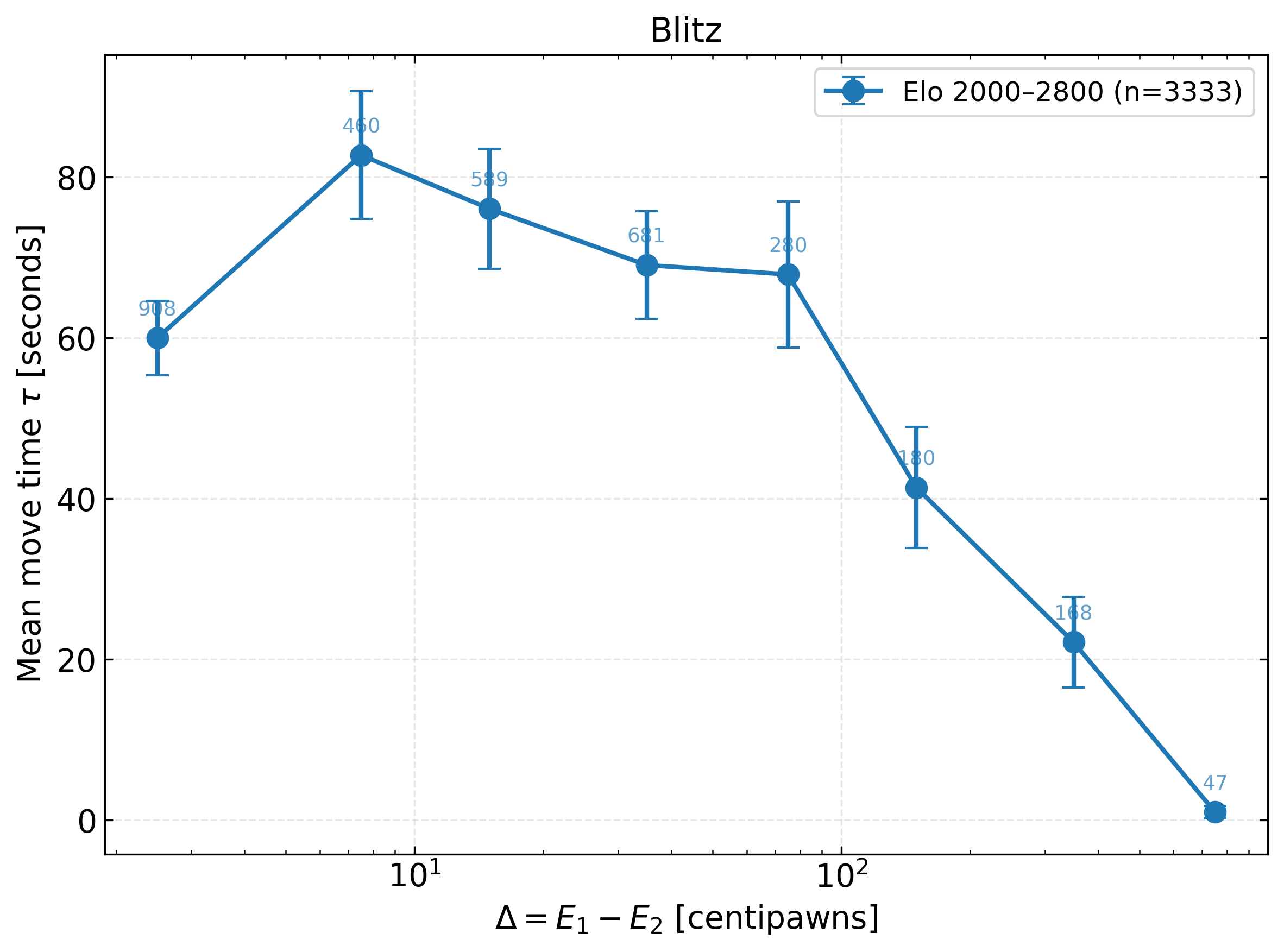}
\caption{Mean thinking time $\langle \tau \rangle$ as a function of the evaluation gap $\Delta$
for Blitz games with Elo $\geq 2000$.
Data are computed from $1000$ games (total of $\sim 3{,}300$ moves)
extracted from~\cite{LumbraGigabase}.}
\label{fig:tau_vs_delta_emp}
\end{figure}

As shown in Fig.~\ref{fig:tau_vs_delta_emp}, the average thinking time $\langle \tau \rangle$
decreases clearly with increasing $\Delta$.
Large values of $\Delta$ correspond to positions in which the best move
is clearly superior to the alternatives, resulting in faster decisions.
Conversely, small $\Delta$ indicates near-indifference between top moves
and therefore greater ambiguity, which typically leads to longer reflection times.

Deviations from strict monotonicity are visible for very small $\Delta$.
These fluctuations can plausibly be attributed to additional factors,
most notably opening preparation:
opening positions often exhibit small $\Delta$ (many roughly equivalent
moves from the engine’s perspective), yet are played rapidly due to memorized theory.
Despite these effects, the overall decreasing trend provides strong empirical support
for $\Delta$ as a relevant proxy for decision-making complexity.

Importantly, we observe a broad maximum of the thinking time around
$\Delta \approx 10$~cp, suggesting that this scale marks a crossover
between positions perceived as genuinely ambiguous and those where
the best move becomes sufficiently distinct.
This empirical observation supports the choice $\Delta_0 \approx 10$~cp
as a natural threshold at expert level.

\subsection{$S(10)$ as a decision-making complexity proxy}

If the entropy $S(10)$ is a meaningful indicator of the intrinsic
complexity of a position, one naturally expects it to correlate with the time players spend deciding on a move. Intuitively, positions with a large entropy~$S(10)$ (i.e., a broad distribution of
plausible continuations) should require more cognitive effort, and
therefore more time, to resolve.

To test this hypothesis, we analyze the same freely available database of online games as above (Lichess \cite{LumbraGigabase}), with move-level time
stamps. The dataset consists of $1\,000$ games, corresponding to a total of $70\,150$ individual moves. For each position, we compute $S(10)$ using Stockfish at depth~15, and record the time~$\tau$ spent by the player on that move.

Since players of different strengths perceive and handle complexity differently, we account for Elo-dependent effects by adapting the reference threshold~$\Delta_0$ as a function of the player's Elo rating. Specifically, we use
\begin{equation}
\Delta_0 =
\begin{cases}
100, & \text{Elo} < 1400, \\[4pt]
100 - 90\,\dfrac{\text{Elo} - 1400}{1000},
      & 1400 \le \text{Elo} \le 2400, \\[4pt]
10,   & \text{Elo} > 2400.
\end{cases}
\label{eq:delta0}
\end{equation}
This choice reflects the increasing ability of stronger players to
discriminate between close alternatives and to process complex positions more efficiently.

\begin{figure}
\centering
\includegraphics[width=0.45\textwidth]{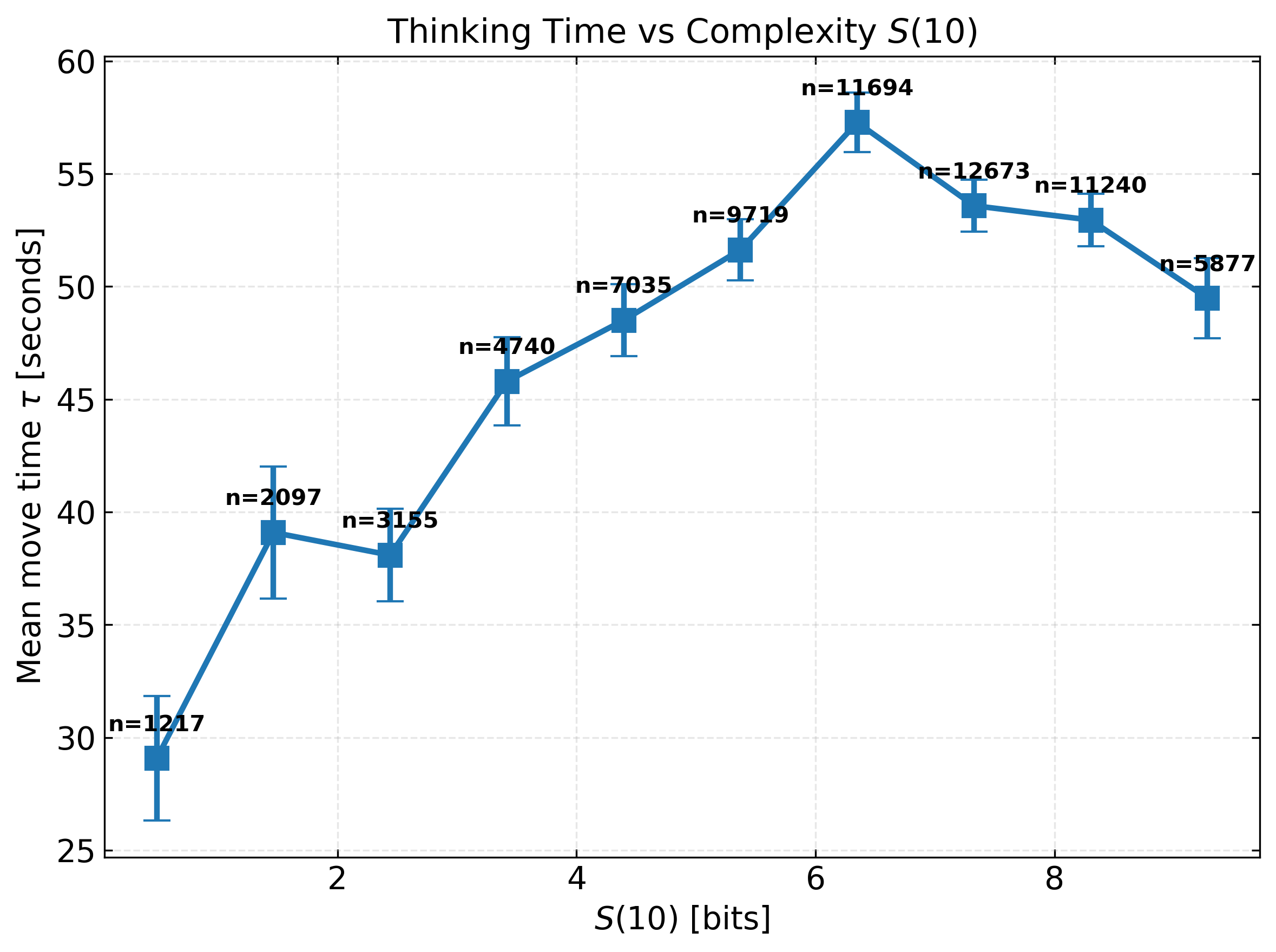}
\caption{Mean thinking time~$\tau$ as a function of the entropy~$S(10)$, computed with the
Elo-dependent threshold~$\Delta_0$ given by
Eq.~\eqref{eq:delta0}. Data are obtained from $1\,000$ Lichess
games ($70\,150$~moves)~\cite{LumbraGigabase}, using Stockfish at
depth~15. Error bars indicate the standard error of the mean.}
\label{fig:tau}
\end{figure}

We then study the dependence of the mean thinking time~$\tau$ on
$S(10)$, as shown in Fig.~\ref{fig:tau}. As expected, we observe an increase of~$\tau$ with~$S(10)$. Deviations from these trends are visible for very small~$\Delta$ or very large~$S(10)$. These fluctuations can be attributed to additional factors not captured by our complexity measures, such as opening preparation: opening positions often exhibit high complexity and small~$\Delta$, yet moves are played rapidly due to memorization.

Despite these effects, the overall trend provides empirical
support for $S(10)$ as a relevant proxy for
decision-making complexity in chess. It captures, in a quantitative manner, the intuitive notion that positions with many competing continuations or poorly differentiated evaluations are cognitively more demanding and require longer deliberation times.

\bibliographystyle{ieeetr}

\bibliography{chess_papers.bib}

\end{document}